\documentclass[sigconf]{acmart}
\AtBeginDocument{%
  }

\setcopyright{rightsretained}
\copyrightyear{2025}
\acmYear{2025}
\acmDOI{10.1145/3706599.3719926}
\acmConference[CHI EA '25]{Extended Abstracts of the CHI Conference on Human Factors in Computing Systems}{April 26-May 1, 2025}{Yokohama, Japan}
\acmBooktitle{Extended Abstracts of the CHI Conference on Human Factors in Computing Systems (CHI EA '25), April 26-May 1, 2025, Yokohama, Japan}
\acmISBN{979-8-4007-1395-8/2025/04}

\begin{document}

\title{Are Cognitive Biases as Important as they Seem for Data Visualization?}

\author{Ali Baigelenov}
\email{abaigele@purdue.edu}
\orcid{0009-0003-6491-1874}
\affiliation{%
  \institution{Purdue University}
  \city{West Lafayette}
  \state{Indiana}
  \country{USA}
}

\author{Prakash Shukla}
\email{shukla37@purdue.edu}
\orcid{0009-0002-7416-1758}
\affiliation{%
  \institution{Purdue University}
  \city{West Lafayette}
  \state{Indiana}
  \country{USA}}

\author{Zixu Zhang}
\email{zhan1575@purdue.edu}
\orcid{0009-0006-5065-2148}
\affiliation{%
  \institution{Purdue University}
  \city{West Lafayette}
  \state{Indiana}
  \country{USA}}

\author{Paul Parsons}
\email{parsonsp@purdue.edu}
\orcid{0000-0002-4179-9686}
\affiliation{%
  \institution{Purdue University}
  \city{West Lafayette}
  \state{Indiana}
  \country{USA}}

\renewcommand{\shortauthors}{Baigelenov et al.}

\begin{abstract}
  Research on cognitive biases and heuristics has become increasingly popular in the visualization literature in recent years. Researchers have studied the effects of  biases on visualization interpretation and subsequent decision-making. While this work is important, we contend that the view on biases has presented human cognitive abilities in an unbalanced manner, placing too much emphasis on the flaws and limitations of human decision-making, and potentially suggesting that it should not be trusted. Several decision researchers have argued that the flip side of biases---i.e., mental shortcuts or heuristics---demonstrate human ingenuity and serve as core markers of adaptive expertise. In this paper, we review the perspectives and sentiments of the visualization community on biases and describe literature arguing for more balanced views of biases and heuristics. We hope this paper will encourage visualization researchers to consider a fuller picture of human cognitive limitations and strategies for making decisions in complex environments.
\end{abstract}



\received{23 January 2025}
\received[revised]{20 February 2025}
\received[accepted]{20 February 2025}

\maketitle

\section{Introduction}
Human decision-making has been an important area of research for several decades. While perfect rationality was long upheld as the ideal, human decision-making has been consistently shown to be sub-optimal by the researchers since the 1950s (e.g., Simon's bounded rationality \cite{simon_behavioral_1955}). In their seminal work, Kahneman and Tversky \cite{kahneman_judgment_1982} argued that people often resort to mental shortcuts (which they referred to as heuristics) when making decisions, especially in situations involving uncertainty. While these heuristics are often useful as they reduce cognitive load, in some cases they lead to cognitive biases, which have been characterized as systematic "errors" in thinking. Many such biases have been documented over the years, and overall the topic has been extremely popular across a wide range of fields. With this popularity, however, there have been critiques that the overall research program has been unbalanced, with researchers advocating for a more nuanced perspective that considers the context and adaptive nature of human decision-making \cite{gigerenzer_how_1991, cohen_can_1981}.

As in other fields dealing with human decision-making, there has been significant interest in biases in the visualization community, where we have seen considerable efforts over the past decade (e.g., \cite{dimara_attraction_2017, wall_left_2021, procopio_impact_2021, ellis_bias_2018}). With this trend, however, the view on decision-making under uncertainty has been somewhat limited, not representing the full range of human capabilities and limitations. Because of this, the visualization community might be missing out on important insights about the adaptive nature of decision-making. In this paper, we provide a preliminary analysis of sentiments towards cognitive biases and heuristics in the visualization literature from the past decade. We draw on literature from psychology and decision making to describe a more balanced perspective of heuristic reasoning that highlights the adaptive and positive aspects of mental shortcuts. Lastly, we present an example to illustrate how a more nuanced view of biases can be utilized in a real world research project involving data visualizations.


\section{Sentiments on Cognitive Biases in the Visualization Literature}
\subsection{Method}
To investigate sentiments on cognitive biases in the visualization literature, we conducted a preliminary analysis using Google Scholar where we searched for literature using a variety of keyword combinations, essentially in the form of ``cognitive bias'' and related terms (e.g., heuristics, fallacies, framing effects) plus (i.e., AND command) ``information visualization'' and related terms (e.g., data visualization, visual analytics). Besides the keyword combinations, our inclusion criteria were as follows:

1) To investigate recent trends and create a manageable starting point for this work, we limited our search to include only literature from the last 10 years (starting from 2014 and onwards, the data collection ended on May 1st of 2024).

2) Included literature must primarily be about visualization (e.g., information visualization, data visualization, visual analytics) \textit{and} cognitive bias (including related concepts such as heuristics, shortcuts, fallacies, and specific biases like anchoring, confirmation, framing effects, and so on). Additionally, both visualizations and cognitive biases have to be prominent concepts in the work (i.e., mentioned in the title and/or abstract). 

Regarding our classification criteria, we have identified a variety of keywords (see supplementary materials for the full list) which were used to determine whether the paper was classified as having negative, neutral, or positive views on cognitive biases and human decision-making. For example, papers were classified as negative if they only used keywords such as \textit{inaccurate, irrational, suboptimal} and \textit{faulty} to describe human decision-making; \textit{imperfections, deviations, mistakes,} and \textit{errors} to describe biases; and \textit{de-biasing, mitigate,} and \textit{alleviate} to frame human decision-making as something that needs to be fixed. Papers in the neutral classification were divided into three categories: neutral-more-negative, neutral-more-positive, and neutral. If a paper mentioned positive sides (in addition to the negative sides) of human decision-making and biases (e.g., mentioned that the underlying mechanisms behind biases are often beneficial, mentioned cases in which "biased" behavior was/is beneficial, and discussed the intricacies of human decision-making through the lens of heuristics and shortcuts rather than only biases) but still adopted a mostly negative sentiment, we classified it as neutral-more-negative. Similarly, if a paper mentioned both positive and negative aspects of human decision-making and biases, but maintained a more positive view throughout the work, we classified it as neutral-more-positive. Then, if a paper maintained a more neutral tone and discussed both negative and positive aspects of the biases and human decision-making roughly equally, we classified it as neutral. Lastly, a paper was classified as positive if it didn't mention any negative aspects (according to the keywords we identified) of biases and human decision-making. 

\subsection{Findings}
This search resulted in 134 papers, of which 22 papers were found to be not relevant in the initial review based on our inclusion criteria; thus, the final list consisted of 112 papers. The final list of 112 papers was collectively coded and analyzed by the first two authors and later discussed with all the authors. The final list of analyzed references (with the codes and comments) is provided as a supplementary material.

Out of the 112 papers analyzed, 71 displayed largely negative sentiments toward human cognitive abilities, emphasizing the flawed nature of the human decision-making and discussing ways of mitigating these perceived flaws. Within the remaining 41 papers, 24 papers presented a somewhat more balanced viewpoint, at least acknowledging that criticisms of the bias-focused work exist or that heuristics can sometimes be useful. 14 of these papers made such acknowledgments (often only superficially or in passing) but remained largely negative throughout the rest of the paper's content. Only three papers (within the remaining 41) adopted a more balanced perspective. None of the papers were purely positive in their view of human cognitive abilities. 

In the sub-sections below, we discuss our findings in relation to our coding strategy described above, categorizing the findings based on literature having (a) predominately negative sentiments, (b) negative sentiments while acknowledging positive aspects, and (c) more neutral or positive sentiments towards human decision making under uncertainty. It is important to note that while we may categorize a paper as negative, we do not view the work itself as negative or flawed. In fact, we are not commenting on the quality of the research at all. We are simply categorizing the sentiments within the papers towards shortcuts, biases, and heuristics.

\subsubsection{Largely negative sentiments}
Many papers adopt a predominantly critical stance towards mental shortcuts, focusing on their detrimental effects on sense-making and decision-making. This sentiment underscores the necessity to detect and mitigate biases in order to enhance decision accuracy and avoid negative consequences. These papers assert that cognitive abilities can routinely disrupt objective reasoning and skew perceptions and decisions, describing \textit{``cognitive biases as predictable deviations from rationality''} \cite{wright_argument_2017} and hypothesizing \textit{``that data visualization users are subject to systematic errors, or cognitive biases, in decision-making under uncertainty''} \cite{wesslen_cognitive_2021}. McNutt et al. \cite{mcnutt_surfacing_2020} note that \textit{``even if visualization designs are not deceptive, our cognitive biases can still cause us to make incorrect or unjustified assumptions about the data.''}. Hilleman et al. \cite{hillemann_role_2015} state that \textit{``to effectively manage information collection and processing while simultaneously avoiding being overwhelmed by too much information, humans unconsciously apply `heuristics' [...] they do not give optimal solutions, just `good enough' solutions that allow humans to save efforts and time but sometimes at the cost of accuracy.''}. Valdez et al. \cite{valdez_framework_2017} highlight a common perspective, namely that shortcuts are evolutionarily outdated---i.e., they helped us avoid threats in our evolutionary past but now cause problems for modern cognitive work: \textit{``Being able to make decisions quickly with limited information and limited resources could make the difference between death by saber-tooth tiger or last-minute escape. Therefore, the human mind is equipped with heuristics that help decision-making with the aim of survival. Today’s world is drastically different! [...] our saber-tooth-fearing minds interfere. Not only is the visual system imperfect but our cognitive system also has its pitfalls. Even when a system provides information perfectly honest, human biases might distort our view of the information and lead to imperfect or outright bad decisions.''}

\subsubsection{Largely negative sentiments while acknowledging potential positives}
Some papers acknowledge potential benefits of heuristic reasoning while still maintaining a focus on the negative aspects (i.e., biases). Law and Basole \cite{ellis_designing_2018} exemplify a style  where a very minor benefit of heuristic reasoning is mentioned (in this case, maintaining analysis flow), while the rest of the paper remains focused on mitigating the negative aspects: \textit{``Unconscious shortcuts are often applied in making these decisions, letting heuristics drive users’ exploration. While heuristics maintains analysis flow by shielding users from making a conscious effort in every step of data exploration, a biased exploration path might hinder insight generation and lead to confirming hypotheses erroneously.''}. Dimara et al. \cite{dimara_attraction_2017} provide a somewhat rare example of acknowledging the criticisms that have been made of the whole biases research program stating: \textit{``one [potential limitation of the study] stems from a general criticism of cognitive bias research, namely, that heuristics that appear irrational may not be so upon deeper examination.''} However, they do not engage with the criticism or discuss what a deeper examination might look like. In follow up work, Dimara and colleagues \cite{dimara_mitigating_2019} elaborate slightly, noting the following: \textit{``Generally, the concept of cognitive bias is controversial within the field of decision making. Some researchers argue that cognitive biases illustrate irrationality, and others argue that some strategies that lead to biases can be effective in complex problems in the real world. Therefore, InfoVis researchers are encouraged to verify if an erroneous response to a visualization task truly reflects irrationality, rather than a strategy based on alternative interpretation of the task.''} While the authors acknowledge the controversial nature of biases, they do not engage with the controversy or discuss how researchers could verify whether an apparent bias truly reflects irrationality. 

\subsubsection{More neutral or slightly positive sentiments} A small number of papers adopt a more neutral sentiment, or at least acknowledge a more nuanced perspective, of heuristic reasoning. They acknowledge the efficiency and practical necessity of heuristics in decision-making, particularly under conditions of uncertainty and limited information. They argue that biases are not always indicative of faulty reasoning, but can also be indicative of adaptive reasoning strategies that, when used correctly, can improve decision-making processes, especially in complex and dynamic settings. For instance, Cottam and Blaha \cite{ellis_bias_2018} suggest that users' goals and intentions shape what they focus on when looking at visualizations to \textit{``keep the information content tractable for human memory and reasoning''}, later stating  \textit{``we disagree that bias only produces unfair outcomes''}, exemplifying a somewhat neutral sentiment. Clearly neutral or positive sentiments were hard to find in core visualization venues (e.g., VIS, CHI, TVCG, EuroVIS), and were more likely to be found in adjacent venues that were still focusing on visualization and cognition. For instance, when discussing lessons from cognitive science for visualizing battlefield information, Geuss et al. from the US Army \cite{geuss_visualizing_2020} suggested that  \textit{``it may be informative to use more behaviorally focused tasks in simulated environments that may better approximate real-world decision-making. Overall, this perspective on decision-making as plagued by cognitive bias suggests a possible opportunity to explore tailored information visualizations as a means to reduce biases in decision-making.''} Booth et al. \cite{booth_representation_2018} theorize heuristic use to explain decision-making from charts, saying \textit{``calculating an accurate risk value based from a line chart is impossible for most people, therefore we must assume that people apply a specific heuristic when evaluating and comparing risk with line charts. Based on this assumption, we theorise that an extrapolation heuristic could explain why changes to chart ratios influence decision-making''} While these examples do not adopt fully positive sentiments towards heuristic reasoning---we did not find any such examples---they at least acknowledge the potential value in a more balanced way.

\section{Naturalistic Decision-Making and Adaptive Expertise}
Researchers in several domains have investigated the reliability of human decision-making in a variety of settings and contexts. While human decisions are often ingenious, adaptive, and effective, they have often been perceived as unreliable and prone to systematic errors (i.e., cognitive biases). Even though these perceptions are true to some degree, the extent to which human decision-making is unreliable can be overstated and the effects of cognitive biases may not be as drastic as they seem when examined outside of the laboratory. In this section we discuss some of the supporting literature to our more balanced view of cognitive biases and human decision-making.

\subsection{Heuristics}
Decision researchers initially upheld perfect rationality as the ideal for human decision making (e.g., see \cite{klein_streetlights_2009}). Kahneman and Tversky, in a series of experiments in their seminal works (e.g., \cite{tversky_availability_1973, kahneman_judgment_1982}) argued against classical decision theory and demonstrated that humans often make decisions that can be statistically "incorrect" or "sub-optimal". When making decisions, especially in complex scenarios and under uncertainty, humans do not necessarily choose what is logical or statistically "correct", but rather employ a set of approximations or "rules of thumb", commonly referred to as heuristics. 

Even before Kahneman and Tversky's seminal work, researchers suggested that humans do not always strive for optimal or perfect and rather often are satisfied with "good enough" (e.g., see Simon's satisficing \cite{simon_models_1996, simon_behavioral_1955}). Humans often employ heuristics, and they are generally effective, but how "good" is "good enough"? Over the decades of research, researchers advanced the narrative that heuristics should not be trusted. Even though the positive sides of heuristics are often mentioned, the experiments run by decision-making researchers have often been designed in ways that can encourage humans to make mistakes (i.e., show the heuristics in a negative way) (e.g., \cite{gigerenzer_homo_2009}). Since Kahneman and Tversky's series of works in the early 1970s, the term heuristics became somewhat of a synonym to biases, where these two concepts are practically always used together, and often seen as interchangeable. This phenomenon is rather interesting, because Kahneman and Tversky were opposing and criticizing classical decision theory, not the decision makers (i.e., humans) themselves \cite{klein_snapshots_2022}. Nor were they necessarily saying that heuristics are bad, but rather that humans \textit{use} heuristics and do not only follow laws of logic, probability, and Bayesian statistics. 

Heuristics have been demonstrated to sometimes outperform more complex decision-making procedures (e.g., \cite{gigerenzer_rationality_2008}). Gigerenzer has continuously argued (e.g., \cite{gigerenzer_heuristic_2011, gigerenzer_homo_2009, gigerenzer_reasoning_1996, gigerenzer_smart_2022}) that heuristics are accurate and effective under uncertainty, referring to them as "fast and frugal" means of reasoning. Gigerenzer has also argued \cite{gigerenzer_homo_2009} that the use of heuristics can be seen as a form of humans making bets (or speculations \cite{klein_snapshots_2022}) in their decision-making process. In complex cases involving uncertainty, involving multiple factors and unknowns, following rational "norms" in the forms of laws of logic, probability, or Bayesian statistics are impossible, and heuristics become invaluable and irreplaceable. Conveniently (or rather ironically), such complex cases are difficult (if not outright impossible) to create in laboratory settings, where the majority of the heuristics and biases research is done. 

\subsection{Models of Rationality}
The inductive nature of heuristics (i.e., "bets" and "speculations") raises a question of which type of rationality these heuristics should be compared to. Heuristics and biases researchers often compare human performance to some baseline or "optimal" performance. As cognitive biases are often referred to as "deviations", the implication is that biased thinking is deviating from an "optimal" or "desired" norm. But what is that "desired" norm? Is there a behavior that is considered "rational"? Rationality in itself is hard to define \cite{ellis_cognitive_2018} and decision-making researchers have discussed various forms of rationality over the years. 

The rationalist paradigm often assumes that deviations from a normative model indicate irrationality in decision-makers \cite{cohen_three_1993}. However, this approach neglects the possibility that the normative model itself may be inappropriate for the specific decision task or context. A more balanced perspective considers that decision-makers' views of their tasks might be flawed, but classical decision theory doesn't always ensure correct decisions, only aiming for the best bet under given conditions. The naturalistic decision-making perspective suggests that decision-making should be sensitive to the environmental constraints and the decision maker's perceptions, which classical theory often fails to adequately include \cite{lipshitz_converging_1993}. Therefore, rather than assuming the decision-maker is wrong, it may be more fruitful to re-evaluate the normative model's assumptions and applicability.

Gigerenzer \cite{gigerenzer_homo_2009, gigerenzer_rationality_2008} introduced a different form of rationality, \textit{ecological rationality}, and argued that the inductive nature of heuristics (i.e., "bets" and "speculations") makes the arguments whether heuristics are "good" or "bad" largely irrelevant, as their effectiveness is significantly affected by the environment. In a sense, the degree to which the heuristics are accurate depends on the context. In this perspective, heuristics are fast and effective ways to deal with the everyday decisions in the real world rather than something that is irrational and error-prone.

\subsection{A More Balanced View of Heuristics and Biases}

The heuristics and biases research program has elevated the judgment and decision-making (JDM) field within psychology by pinpointing systematic errors in human judgment and demonstrating how these errors can be predictably influenced \cite{doherty_laboratory_1993, klein_streetlights_2009}. This includes the understanding that biases are not merely random mistakes, but rather reveal fundamental aspects of how our minds operate. This research aids in understanding how people think by exploring reasoning strategies, with experiments illustrating how these strategies function. Findings from this research program can certainly inform the design of decision support systems and interventions aimed at mitigating the impact of biases in real-world settings, including in visualization and visual analytic contexts. Despite some reservations about terminology, the presence of biases, or systematic deviations from normative standards, is generally accepted as a real and important phenomenon \cite{doherty_laboratory_1993}. These biases are viewed as limitations in reasoning strategies that can lead to errors, even if the strategies themselves are generally effective and hold value. This underscores that acknowledging decision biases means recognizing the limitations in our reasoning strategies, while also understanding that these strategies are often effective in complex, natural settings involving uncertainty and limited information.

One of our aims in this work is to raise awareness within the visualization community about the rich discussions, and variety of perspectives, relating to human cognition and heuristic reasoning that exist in several sub-fields of the decision making literature. Rather than focusing almost exclusively on mitigating biases, researchers may benefit from adopting alternative perspectives that recognize the value of heuristics in complex environments. This is particularly important as visualization researchers are becoming more focused on the role of context and the environment in data analytics and visualization (e.g., \cite{thomas_situated_2018, bressa_whats_2021, elmqvist_data_2023}). Recognizing that heuristics can be ecologically rational in action---while appearing to be biases in a lab setting---is critically important for the study of situated visualization initiatives. In the next section, we aim to demonstrate the benefits of this view with a scenario.

\subsection{Methodological Issues}
The cognitive biases research program has faced substantial criticism regarding the ecological validity and limitations inherent in its reliance on laboratory experiments \cite{beach_why_1993, klein_streetlights_2009}. A primary concern is that these experimental conditions often fail to mirror the complexities and dynamics of real-world decision-making environments. Tasks frequently involve simplified, context-limited scenarios that do not adequately represent the challenges faced by experts operating in their natural domains \cite{orasanu_reinvention_1993}. This discrepancy raises questions about whether findings from experiments can be reliably generalized to real-world settings, where decisions are influenced by a multitude of factors such as expertise, tacit knowledge, team cognition, established procedures, task-specific knowledge, and environmental cues. The artificiality of these experiments can lead to an overestimation of the prevalence and impact of decision biases, as individuals may employ different strategies and reasoning processes when confronted with real-world stakes and complexities \cite{klein_snapshots_2022}.

Critics argue that the emphasis on identifying deviations from normative models overlooks the adaptive nature of human cognition \cite{cohen_three_1993}. In real-world settings, individuals often develop heuristics and strategies that are "ecologically rational," meaning they are well-suited to the structure of the environment \cite{gigerenzer_heuristic_2011}. These strategies may deviate from formal norms but still lead to effective outcomes, given the constraints of limited time, information, and cognitive resources. Furthermore, laboratory experiments often utilize non-random stimuli and subject selection, potentially biasing results towards the detection of suboptimal processes. The focus on statistical significance, rather than meaningful effect sizes, further compounds the issue, as statistically significant biases may have negligible practical consequences in real-world domains \cite{cohen_naturalistic_1993}.


By expanding research methods to include the study of decision-making in realistic, dynamic, and complex environments, a more comprehensive understanding of human reasoning and its limitations can be achieved. This involves adopting methodologies that directly focus on decision processes and their real-world outcomes, allowing for a more nuanced evaluation of decision errors and the factors that contribute to them.

\section{Scenario: Studying Data Visualization Use in a Complex Sociotechnical System}

To illustrate the implications of adopting an approach that is not focused on mitigating biases, we describe a project we have been working on and how the perspective taken on decision-making influenced the outcomes. We have been working on a NASA-funded project whose goal is to develop knowledge and methods for space habitation beyond low-earth orbit \cite{murali_krishnan_habsim-hms_2024}. Our responsibility on the project is in regard to human decision-making and data visualization. Our mandate was open-ended, essentially to develop knowledge that could be useful for future missions. Our first step was to study the current practice of decision-making with the closest analog to a distant space habitat---the International Space Station (ISS). Mission control for the ISS consists of a large team of flight controllers and other personnel, including a team that works in the front room monitoring vast amounts of data down-linked from the station. These controllers typically sit in front of 4-6 computer screens that display vast amounts of data, including dozens of visualizations (most typically line charts). We decided to focus on understanding how current mission control operators made decisions with data-driven displays in the face of change and uncertainty. One approach to doing this would have been to focus on bias(es) and make recommendations for how to mitigate them with interactive visualization techniques. Indeed, this was a direction suggested by members of the advisory board for the project. Following this approach, we likely could have identified one or more biases at play, and set up experimental conditions to study them. If the heuristics and biases view was our dominant perspective on decision-making, we may have adopted this approach.  

Instead of starting our investigations through the lens of bias---i.e., assuming decision-making is flawed, and looking for shortcomings that could be mitigated---we adopted an alternative lens. This view suggests that experienced decision-makers are adaptive experts, using mental shortcuts to act quickly and respond to dynamic, high-pressure situations. We decided to study the decision making strategies of mission control operators, identifying how they operate and response to anomalies (see \cite{zhang_designing_2023} for more details). During our study, we found that flight controllers never seem to rely too heavily on an individual data point or visualization. Instead, they corroborate their hypotheses across a variety of sources, including their systems knowledge, procedures and schematics, and especially others working in mission control. The chances that a visualization or set of visualizations would lead to an erroneous or biased decision in this context is very small. Furthermore, several participants told us the data per se is often not terribly important (keep in mind, this is in a context where they are literally flying a space station and have thousands of data points being down-linked every second). Rather, participants told a story where they had to be keenly aware of the constraints and pressures of a variety of systems at play, drawing on their training and knowledge of current and ongoing dynamics, all while communicating with potentially dozens of other people working within mission control. The vast array of data points and charts on their displays were only one piece of the decision making puzzle---mission control operators monitored data streams, but also spoke with other operators, confirmed data with astronauts on the station, checked their own training and systems knowledge, and engaged in a variety of procedures and protocols for flagging potentially worrisome trends in the data. Frequently, some kind of hunch or gut feeling would lead to the investigation of a problem or potential problem, exemplifying the use of heuristic reasoning. Such strategies can then be used to stimulate ideas for designing new visualizations and interactive interfaces (e.g., see some our work regarding visualizing potential anomalies with boundary lines \cite{zhang_visual_2022}). If the same decision making strategies were isolated and studied in a lab, they may have been characterized negatively as exemplifying biased reasoning. However, in the world of practice, where heuristics may exemplify adaptive rationality within a particular environment, such biases may rarely or never manifest in the same ways.

It is easy in retrospect to claim bias as a factor in failures and accidents, but many reports attribute `human error' as a cause without doing due diligence to uncover the real causes (which are often in spite of human adaptation rather than because of human limitations) \cite{dekker_field_2017}. Our investigations uncovered a wide variety of examples where flight controllers were faced with anomalies and were able to adapt and generate solutions. Dozens of unexpected, non-routine situations that flight controllers handled, if biases were a major cause for concern, we would expect to see more negative effects. However, despite the limited attention given to biases in space mission control, we see a high degree of success in decision-making and anomaly response. The ISS has been in continuous flight for more than 20 years, and cognitive biases have not been a major cause for concern thus far.

Complex systems are brittle and regularly at risk of failing \cite{woods_joint_2006}. It is often the humans in the systems that are the sources of resilience \cite{hollnagel_coping_2012}. Without human adaptability and expertise, these systems would fail regularly. So, do we put more attention on the limitations that might exist (e.g., biases) or on the expertise and adaptation (e.g., heuristics as ecological adaptations)? The biases lens can tell us useful things, but with too much emphasis we will not learn about the way decision makers adapt and work around constraints and limitations imposed by the environment. While we acknowledge that this scenario is somewhat speculative and does not offer empirical validation, we believe it still helps to illustrate the argument we are trying to make as it is based on a real world research project.

We believe the visualization research community would benefit from adopting a view on cognitive biases and decision-making that is more inclusive of the variety of perspectives on heuristics and biases that exist in the broader literature. As visualization researchers become increasingly interested in situated visualization, ubiquitous visualization, and other forms of context-dependent research and design, awareness of such perspectives can be valuable, as researchers will need to make decisions akin to the ones we employed in the scenario above. We could have entered the project with a biases mindset, identified tasks and designed experiments, and we likely would have identified biases relevant to the decision making that takes place in mission control. However, we decided to focus on how humans make judgments and adapt to the context in which they find themselves. In doing so, we discovered how data visualizations were used as only one piece of the decision making puzzle, and their features were considered along with many other procedures and protocols within the context of mission control. Our work has led us to consider how might we provide alternative visualization tools to help operators develop hypotheses, explore alternative explanations, and envision where data streams may be going in the future, rather than thinking about how can we use visualization tools to mitigate cognitive biases.

Our scenario of course exemplifies only one domain and context of visualization use---e.g., a high pressure scenario involving uncertainty, dynamism, expertise, and critical decisions that need to be made in a team environment. This can be contrasted with more casual scenarios where users interpret a visualization in a news article or business report, for example. While it may be that biases are more relevant or important in some contexts over others---and in some contexts researchers \textit{should} focus specifically on bias mitigation---we believe it would still benefit the visualization community to have a more inclusive and robust theoretical and methodological perspectives on heuristic reasoning and decision making.

\section{Conclusion}

The study of cognitive biases has played a significant role in visualization research, shaping how we understand human decision-making in data-driven contexts. However, as we have argued, this focus has often presented an overly negative view of human cognition, emphasizing errors and deviations rather than the adaptive strengths of heuristic reasoning. Through our analysis of the past decade of visualization literature, and our real-world example, we have demonstrated that heuristic-based decision-making is not inherently flawed but often represents ecologically rational strategies that enable users to navigate complexity effectively. For visualization researchers and designers, this perspective suggests a need to rethink how cognitive biases are framed in research and design. Rather than treating biases primarily as obstacles to be mitigated, we advocate for a more balanced approach that acknowledges the functional role of heuristics in real-world decision-making. This shift has practical implications: visualization tools should not only aim to correct for biases but should also be designed to support and enhance the adaptive strategies that users employ in complex environments. By broadening the theoretical foundation of visualization research to incorporate perspectives from naturalistic decision-making and adaptive expertise, we can move beyond the limitations of traditional bias-focused studies. Future research should explore how visualization systems can be designed to align with users’ heuristic reasoning processes, enabling more effective sense-making, hypothesis generation, and decision support. In doing so, we can develop visualization techniques that are not only more aligned with human cognition but also more useful in real-world settings.


\bibliographystyle{ACM-Reference-Format}
\bibliography{references}


\end{document}